\documentclass[aps,twocolumn,showpacs,preprintnumbers,nofootinbib,prd,superscriptaddress,groupedaddress,10pt]{revtex4-1}

\makeatletter
\def\l@subsubsection#1#2{}
\def\l@subsubsubsection#1#2{}
\makeatother

\usepackage{graphicx,amssymb,amsmath,amsthm,amsfonts,epsfig,epsf}
\usepackage[usenames]{color}
\usepackage{epstopdf}
\usepackage[caption=false]{subfig}

\usepackage{bm}
\usepackage{dcolumn}
\usepackage{latexsym}
\usepackage{rotating}
\usepackage{longtable}

\usepackage{enumerate}
\usepackage[inline]{enumitem}
\usepackage{tensor,multirow}
\usepackage{url}
\usepackage[linktocpage]{hyperref}

\usepackage{physics}

\begin{document}
	
	\title{Ring formation from black hole superradiance through repeated particle production on bound orbits}
	
	\author{Zhen-Hong Lyu$^{1,2}$}
	\email{lyuzhenhong@itp.ac.cn}
	
	\author{Rong-Gen Cai$^{3}$}
	\email{caironggen@nbu.edu.cn}
	
	\author{Zong-Kuan Guo$^{1,2,4}$}
	\email{guozk@itp.ac.cn}
	
	\author{Jian-Feng He$^{1,2}$}
	\email{hejianfeng@itp.ac.cn}
	
	\author{Jing Liu$^{5,6}$}
	\email{liujing@ucas.ac.cn}
	
	\affiliation{$^1$Institute of Theoretical Physics, Chinese Academy of Sciences, Beijing 100190, China}
	\affiliation{$^2$School of Physical Sciences, University of Chinese Academy of Sciences, Beijing 100049, China}
	\affiliation{$^3$Institute of Fundamental Physics and Quantum Technology, \& School of Physical Science and Technology, Ningbo University, Ningbo, 315211, China}
	\affiliation{$^4$School of Fundamental Physics and Mathematical Sciences, Hangzhou Institute for Advanced Study, University of Chinese Academy of Sciences (HIAS-UCAS), Hangzhou 310024, China}
	\affiliation{$^5$International Centre for Theoretical Physics Asia-Pacific, University of Chinese Academy of Sciences, Beijing 100190, China}
	\affiliation{$^6$Taiji Laboratory for Gravitational Wave Universe (Beijing/Hangzhou), University of Chinese Academy of Sciences, Beijing 100049, China}

	\begin{abstract}
		Ultralight bosonic fields around a rotating black hole can extract energy and angular momentum through the superradiant instability and form a dense cloud. We investigate the scenario involving two scalar fields, $\phi$ and $\chi$, with a coupling term $\frac{1}{2}\lambda\phi\chi^2$, which is motivated by the multiple-axion framework. The ultralight scalar $\phi$ forms a cloud that efficiently produces $\chi$ particles nonperturbatively via parametric resonance, with a large mass hierarchy, $\mu_\chi \gg \mu_\phi$. Rather than escaping the system as investigated by previous studies, these $\chi$ particles remain bound, orbiting the black hole. Moreover, the particle production occurs primarily at the peak of the cloud's profile, allowing $\chi$ particles in quasicircular orbits to pass repeatedly through resonant regions, leading to cumulative amplification. This selective process naturally forms a dense ring of $\chi$ particles, with a mass ratio to the cloud fixed by $(\mu_\phi/\mu_\chi)^2$. Our findings reveal a novel mechanism for generating bound-state particles via parametric resonance, which also impacts the evolution of the cloud. This process can be probed through its imprint on binary dynamics and its gravitational-wave signatures.
		
	\end{abstract}
	
	\maketitle
	
	\section{Introduction}

	The interaction between a rotating black hole~(BH) and an ultralight bosonic field can trigger a powerful instability known as superradiance~\cite{zeldo1,Damour:1976kh,Teukolsky:1974yv,Press:1972zz}.
	(see, for a comprehensive review,~\cite{Brito:2015oca}). This process extracts rotational energy and angular momentum from the black hole, forming an exponentially growing boson cloud with a hydrogenlike structure, known as a ``gravitational atom."
	The efficiency of this instability depends on the ratio of the black hole’s gravitational radius to the Compton wavelength of the bosons, which is maximized when the gravitational fine structure constant~\cite{Baumann:2019eav},
	\begin{equation}
		\alpha\equiv GM_{\rm BH}\mu=0.08 \left( \frac{\mu}{10^{-12}\mathrm{eV}} \right)  \left( \frac{M}{10 M_{\odot}} \right)\,,
	\end{equation}
	is $\mathcal{O}(0.1)$ (in natural units, $\hbar=c=1$). For astrophysical black holes with masses ranging from $\sim 5M_\odot$ to $\sim 10^{10}M_\odot$, this condition makes the superradiant instability an ideal probe for bosons in the ultralight mass range $10^{-20}$ to $10^{-10}$ eV.
	
	This physical insight has transformed astrophysical black holes into laboratories for probing new physics, offering a unique discovery channel for weakly interacting, ultralight bosons predicted by theories beyond the Standard Model, most notably axionlike particles~\cite{Arvanitaki:2009fg,Arvanitaki:2010sy}. These particles are particularly compelling candidates in this context, as their naturally small mass can be protected by an underlying shift symmetry, broken only by nonperturbative effects~\cite{Bauer:2020lowenergy}. Furthermore, these particles are well-motivated dark matter candidates~\cite{Hui:2016ltb}. The formation of an axion cloud around a Kerr black hole via superradiance has been a subject of intense research. This includes foundational theoretical investigations~\cite{Shlapentokh-Rothman:2013ysa,Moschidis:2016wew,Ficarra:2018rfu,Yang:2023vwm,Luo:2024gqo}, numerical studies~\cite{Dolan:2012yt,Okawa:2014nda} and diverse phenomenological studies. The phenomenological explorations cover topics from constraining axion properties with black hole observations~\cite{britoGravitationalWaveSearches2017,Stott:2018opm,Fernandez:2019qbj,chengConstrainsAxionlikeParticle2023,Saha:2022hcd,Bai:2025yxm} to identifying unique gravitational wave signatures~\cite{Ghosh:2018gaw,Guo:2022mpr,liuGravitationalLaserStimulated2024} and assessing the impact of stellar companions on the cloud~\cite{Baumann:2018vus,baumannGravitationalColliderPhysics2020,Zhang:2019eid,Takahashi:2021yhy,Cao:2023fyv,Tong:2022bbl,Tomaselli:2024dbw}. Similar superradiant phenomena for vector and tensor fields have also been extensively explored~\cite{Pani:2012vp,Baryakhtar:2017ngi,Cardoso:2018tly,East:2017ovw,Brito:2020lup}. 
	
	The majority of these theoretical studies have focused on the dynamics of a single, noninteracting field. When accounting for self-interactions or couplings to other fields, the superradiant dynamics can be dramatically changed. For instance, self-interactions may lead to scalar emission, level mixing, and the bosenova explosions phenomena~\cite{Arvanitaki:2010sy,Gruzinov:2016hcq,Baryakhtar:2020gao,Xie:2025npy}. On the other hand, interactions between the superradiant field and other particle species can be equally consequential, leading to complex dynamics and rich phenomenology~\cite{Fukuda:2019ewf}. Generally, such nonlinearity can trigger resonant production of new particles directly from the superradiant condensate. This process drains energy from the cloud by creating a large number of these new particles, thereby suppressing or quenching the superradiant growth~\cite{Fukuda:2019ewf}. Prominent examples of this mechanism include the coupling to photons, which can source powerful, observable bursts of electromagnetic radiation~\cite{rosaStimulatedAxionDecay2018,Ikeda:2018nhb,Sen:2018cjt,boskovicAxionicInstabilitiesNew2019,Spieksma:2023vwl}, and the coupling to neutrinos, which can generate significant fluxes of high-energy neutrinos~\cite{Chen:2023vkq}.
	
	In this work, we investigate an interacting two-axion system ($\phi, \chi$) around a rotating black hole, where the lighter superradiant axionlike particle $\phi$ couples to the heavier one $\chi$ via a trilinear coupling, $\frac{1}{2}\lambda\phi\chi^2$. Such an interaction drives a powerful, nonperturbative production of $\chi$ particles beyond the kinematic limitations of perturbative decay. We demonstrate a novel ring-formation mechanism that stands in contrast to scenarios where produced particles typically escape the system. Our analysis highlights two key insights. First, particle production commences at the peak of the cloud's profile, which naturally selects for gravitationally trapped particles on quasicircular orbits. Second, unlike escaping particles that are amplified only once, these trapped particles repeatedly traverse the resonant regions of the cloud. This leads to cumulative amplification, causing their population to grow exponentially and eventually form a stable ring of $\chi$ particles. The existence of multiple, coupled axion fields—an ``axiverse"—is well-motivated by fundamental theories like string theory, where a large number of axions with a vast range of masses naturally arise from the compactification of higher-dimensional gauge fields~\cite{Bauer:2020lowenergy,PeterSvrcek_2006,Cicoli:2012sz,Arvanitaki:2009fg}.
	
	The rest of this paper is organized as follows.
	In Sec.~\ref{sec:production}, we review the superradiant instability and detail the resonant production of $\chi$ particles from the oscillating $\phi$ cloud. In Sec.~\ref{sec:ring}, we analyze the postproduction dynamics, employing a WKB approximation to model particle trajectories and demonstrating how cumulative amplification leads to the formation of the particle ring. We also discuss the backreaction of the ring on the cloud to determine the final, saturated state of the system. We conclude in Sec.~\ref{sec:Discussion}. We will use a metric with ``mostly plus" signature $(-,+,+,+)$ and work in natural units with $\hbar=c=1$.
	
	\section{Resonant Particle production from superradiant clouds}~\label{sec:production}
    
	\subsection{Black hole superradiance}
	We start by reviewing the superradiant instability for massive scalar fields. The equation of motion for a free massive real scalar field $\phi$ with mass $\mu$ around a Kerr black hole is
	\begin{equation}
		\Box\phi=\mu^2\phi\,.
	\end{equation}
	If the gravitational coupling is weak, i.e., $\alpha\ll 1$, the wave equation can be solved analytically through the ansatz $\phi\sim \Re\int\dd\omega\sum_{\ell m} e^{-i\omega t+im\varphi}R_{\omega\ell m}(r)S_{\omega\ell m}(\theta)$, where $\omega$ is the eigenfrequency and $\ell,m$ are the angular quantum numbers.

    The quasibound state solutions have a discrete spectrum of complex eigenfrequencies,  $\omega=\omega_R+i\omega_I$. The superradiance instability arises due to the presence of growing modes for $\phi$, which can be identified as the positive imaginary eigenvalues, i.e., $\omega_I>0$. This also implies the superradiance condition
	\begin{equation}
		m\Omega_{\rm H}>\omega_R\,,
	\end{equation}
	where $\Omega_{\rm H}=\tilde{a}/2r_+$ is the angular velocity of the black hole’s event horizon, with $\tilde{a}$ the black hole's dimensionless spin and $r_+=GM_{\rm BH}(1+\sqrt{1-\tilde{a}^2})$ the horizon radius.
	
	The energy levels represent hydrogenlike features at leading order of $\alpha$~\cite{detweilerKleinGordonEquationRotating1980,dolanInstabilityMassiveKleinGordon2007a,Baumann:2019eav}:
	\begin{equation}
		\omega_{R,n}=\mu\left( 1-\frac{\alpha^2}{2n^2} \right)+{\cal O}(\alpha^4)\,.
	\end{equation}
	
	The fastest growing mode $(n\ell m)=(211)$ dominates the scalar field evolution, with the growth rate~\cite{detweilerKleinGordonEquationRotating1980,dolanInstabilityMassiveKleinGordon2007a,Baumann:2019eav}:
	\begin{equation}\label{eq:211}
		\Gamma_{211}\simeq \frac{1}{48}(\tilde{a}-4\alpha)\alpha^8\mu\,.
	\end{equation}
	
	The exponential growth of this dominant mode results in the scalar field extracting energy and angular momentum from the black hole, forming a nonspherical, rotating condensate. In the nonrelativistic limit, the wave function of the $(211)$ state is well approximated by~\cite{Chen:2023vkq}
	\begin{equation}
		\phi(t, r, \theta, \varphi) = \Phi_0(t) \left(\frac{r}{2r_0}\right) e^{1 - r/2r_0}\sin \theta \cos(\varphi - \mu_{\phi} t)\,.\label{211profile}
	\end{equation}
	
	Here, $\Phi_0(t)\sim e^{\Gamma_{211}t}$ represents the time-dependent amplitude of the field, which grows exponentially at the superradiant rate $\Gamma_{211}$. The characteristic size of the cloud is set by the gravitational Bohr radius,
	\begin{equation}
		r_0=\frac{1}{\alpha\mu}\simeq 2\times 10^{6}\left(\frac{0.1}{\alpha}\right)\left(\frac{10^{-12}\mathrm{eV}}{\mu}\right)\mathrm{m}\,.
	\end{equation}
	
	The Newtonian approximation for these wave functions is valid in the region where $r\gg r_{g}=GM$. Since the Bohr radius is on the order of $r_{0}=r_{g}/\alpha^2$, the approximation is well justified for small $\alpha$, as the vast majority of the cloud resides far from the black hole. A key feature of this state is that its density profile peaks at a radius $r = 2r_0$ within the equatorial plane ($\theta = \pi/2$).
	
	When the angular momentum of the black hole is extracted to $m\Omega_{\rm H}=\omega_R$, i.e., the superradiance condition is no longer met, the scalar cloud saturates. The energy density of the cloud is approximately proportional to the potential energy, $\rho\sim \mu^2\phi^2$, leading to the total mass of the cloud at saturation~\cite{Chen:2022kzv},
	\begin{equation}\label{cloudmass}
		M_{\rm cloud}=\int\dd[3]\mathbf{r} \rho(\mathbf{r})\simeq \frac{186\Phi_0^2}{\alpha^3\mu}\,.
	\end{equation}
	It has been investigated that the maximal value of $M_{\rm cloud}$ approximates about 10\% of the black hole mass~\cite{britoBlackHolesParticle2015,Herdeiro:2021znw}.
	
	\subsection{Interacting two-axion superradiant system}
	We consider a system that contains another scalar $\chi$, which is coupled with $\phi$ through the interaction term $V_{\mathrm{int}}$. The Lagrangian reads\footnote{In the small-field regime relevant for superradiant growth, expanding the full periodic potential, for instance, $V_\phi=\mu_{\phi}^2f^2[1-\cos\left(\phi/f\right)]$, where $f$ is the axion decay constant, yields the quadratic mass term as well as quartic self-interaction at leading order. Similarly, the effective polynomial interaction term $V_{\mathrm{int}}$ can be derived from a full potential within an effective field theory framework. Although the quartic self-interaction is important, this work focuses on exploring the distinct phenomenology from couplings between different fields. We therefore neglect self-interaction effects to focus on this new dynamic.}
	\begin{equation}
		-\mathcal{L} = \frac{1}{2}(\partial \phi)^2 + \frac{1}{2}(\partial \chi)^2 + \frac{1}{2}\mu_{\phi}^2\phi^2 + \frac{1}{2}\mu_{\chi}^2\chi^2 + V_{\mathrm{int}}\,,
	\end{equation}
	where $\mu_{\phi}$ and $\mu_{\chi}$ are the masses of the two scalar fields. In this work, we focus on a scenario with a significant mass hierarchy, $\mu_{\phi}\ll \mu_{\chi}$, which can be motivated by the ``axiverse" scenarios.\footnote{We expect the mass hierarchy would not be extremely large, otherwise we can integrate the heavy field and introduce the effective self-interaction of $\phi$, see Sec.3.4 of Ref.~\cite{Fukuda:2019ewf} for an example of four-point interaction.}  The lighter field, $\phi$, satisfies the superradiant condition, meaning its gravitational fine-structure constant $\alpha=GM_{\rm BH}\mu_\phi$ is ${\cal O}(0.1)$. The mass hierarchy then implies that the corresponding parameter for the heavier field, $\alpha_\chi=GM_{\rm BH}\mu_\chi$ is much larger than one, i.e., $\alpha_{\chi}\gg 1$. Consequently, the superradiant instability for the $\chi$ field is strongly suppressed~\cite{dolanInstabilityMassiveKleinGordon2007a}, and only the $\phi$ field can form a primary cloud via this mechanism.
	
	To explore the dynamics between the two fields, we consider the possible interactions in the potential $V_{\rm int}$. While a general potential can include a variety of renormalizable operators, our analysis will focus exclusively on the trilinear interactions capable of inducing efficient particle production via parametric resonance:
	\begin{equation}
		V_{\mathrm{int}}=\frac{1}{2}\lambda \phi \chi^2\,.
	\end{equation}
	
	We acknowledge that other couplings exist. The quartic interaction $\frac{1}{2}g^2\phi^2\chi^2$ also leads to efficient particle production but yields distinctly different phenomenology, as it drives an escaping flux of accelerated particles rather than forming a stable bound state. We will address this case in the Discussion (Sec.~\ref{sec:Discussion}) with details provided in Appendix~\ref{app:interaction}. Other possible renormalizable operators are shown to be ineffective at producing resonance, as they are either subdominant or kinetically forbidden (see Appendix~\ref{app:interaction} for details).
	
	\subsection{Particle production from the trilinear interaction}
	
	We now analyze the particle production dynamics induced by the trilinear coupling, $V_{\rm int}=\frac{1}{2}\lambda \phi\chi^2$. In this setup, the macroscopic, coherently oscillating $\phi$ cloud acts as a classical background field amplifying fluctuations of the $\chi$ field, leading to exponential particle production. The mechanism of particle production from a coherent scalar field is well known, most notably from studies of reheating in early-universe cosmology, where a homogeneous inflaton field is responsible for matter production~\cite{Boyanovsky:1996sq,Kofman:1997yn,Bassett:2005xm}. Our scenario introduces crucial distinctions from the typical homogeneous and inflaton background: the superradiant $\phi$ cloud is both spatially inhomogeneous, with a distinct toroidal profile, and its amplitude grows exponentially over time. Initially, when the interaction term is negligible, the amplitude of the $\phi$ field, $\Phi_{0}(t)$, grows exponentially through superradiant instability. Once $\Phi_{0}(t)$ reaches a certain threshold, the oscillation amplitude of $\chi$ field can be enlarged exponentially through parametric resonance.
	
	As mentioned before, the characteristic scale of the axion cloud is $r_{0}=(\mu_{\phi}\alpha)^{-1}$. While modes with wavelengths comparable to or larger than this scale might exist, our analysis focuses on the short wavelength modes, $k \gg \mu_{\phi}\alpha$, corresponding to the wavelength of $\chi$ that is significantly smaller than the cloud radius. In this way, one can neglect finite-size effects of the cloud, and a local plane wave approximation of $\chi$ is valid. Besides, since such a wavelength is much smaller than the distance from the central black hole, i.e., $r_0 \gg r_{g}$, we can neglect curvature effects in studying the local $\phi-\chi$ interactions in this region. Furthurmore, since the oscillation timescale $\mu_\phi^{-1}$ is much shorter than that of superradiance growth $\Gamma_{211}^{-1}$, at some fixed time $t_{\star}$, we may approximate the $\phi$ cloud as coherently oscillating classic field with static amplitude $\Phi_{0\star}=\Phi_{0}(t_{\star})$ in Eq.\eqref{211profile}.
	
	Under these approximations, the equation of motion for a local $\chi_k$ mode is given by
	\begin{equation}
		\ddot{\chi}_{k}+\omega_{k}^2(t)\chi_{k}=0\,,
	\end{equation}
	with the time-dependent effective frequency squared
	\begin{equation}
		\omega_{k}^2=k^2+\mu_{\chi}^2+\lambda \Phi_{0\star}\cos(\varphi-\mu_{\phi}t)\,.
	\end{equation}
	
	The $\chi$ field is amplified at the instances when $\omega_{k}^2$ is minimal, at which the nonadiabatic condition
	\begin{equation}
		\abs{\frac{\dot{\omega}_{k}}{\omega_{k}^2}}\gtrsim 1
	\end{equation}
	is triggered. In our setup, this occurs on planes where approximately $\varphi=\left( 2l+1\right)\pi+\mu_{\phi}t$ ($l\in\mathbb{N}$) and $\phi=-\Phi_{0\star}$. Outside these brief instances of nonadiabaticity, the evolution remains largely adiabatic.
	
	For a fixed amplitude $\Phi_{0\star}$, the range of wave numbers $k$ that are significantly amplified can be estimated from the stability analysis of the Mathieu equation. In the regime of interest where the pump field is strong ($\lambda\Phi_{0\star}/\mu_{\phi}^2\gg 1$), the condition for significant amplification is
	\begin{equation}\label{eq:condition}
		k^2+\mu_{\chi}^2\lesssim \lambda \Phi_{0\star}+\frac{\mu_{\phi}}{2}\sqrt{2\lambda \Phi_{0\star}}\,.
	\end{equation}

	Note that the broad resonance begins with $k^2+\mu_{\chi}^2\gtrsim \lambda \Phi_{0\star}$, i.e., $\omega_k^2>0$, due to a minor correction represnented by the second term on the right-hand side, in contrast to tachyonic resonance which requires $\omega_k^2<0$.
	
	This condition dictates the \textit{spatiotemporal onset} of particle production. The process becomes efficient only when the cloud's amplitude $\Phi_0$ reaches $\lambda \Phi_0\simeq \mu_\chi^2$. At the threshold, the first modes to be excited are those with the lowest possible $k$ (while remaining consistent with our local approximation, $k \gg \alpha\mu_\phi$). Crucially, the profile of the superradiant cloud is not uniform; it has a distinct toroidal profile (as described in Eq.~\eqref{211profile}) that peaks at a radius of $r=2r_0$ in the equatorial plane, $\theta=\pi/2$. Therefore, as the overall cloud grows, this peak toroidal region is the \textit{first} location in space to reach the resonant threshold. Particle production thus ignites within this specific ringlike zone. As $\Phi_{0}(t)$ continues to grow, the resonance condition is met for progressively larger momentum and across a wider region, opening up the production of more energetic $\chi$ particles.

	\section{Postproduction of the $\chi$ particle and formation of the particle ring}~\label{sec:ring}
    
	In contrast to many scenarios in the literature where particle production is a dissipative process creating escaping relativistic fluxes~\cite{rosaStimulatedAxionDecay2018,Ikeda:2018nhb,Sen:2018cjt,boskovicAxionicInstabilitiesNew2019,sunFastGravitationalWave2021,Chen:2023vkq}, we explore a qualitatively different outcome. As established in the last subsection, the resonance first produces low-momentum particles. We will demonstrate that a select fraction of these low-momentum $\chi$ particles can be gravitationally trapped and confined to stable, quasicircular orbits within the primary $\phi$ cloud. Unlike all other particles that either escape the system or are quickly accreted, this orbiting population repeatedly traverses the resonant regions. This allows for cumulative amplification, a mechanism unique to these bound trajectories, which causes them to become the dominant component of the final state, a critical feature absent in the escaping case. This leads to the formation of a novel toroidal structure of $\chi$ particles in a nondissipative way for energy redistribution within the superradiant system. To substantiate this picture, we will now derive the classical equations of motion using the WKB analysis within the weak-field limit.

	\subsection{The classical dynamics and the formation of the particle ring} 
	
	The $\chi$ particles are produced as localized wave packets from the resonance region of the $\phi$ cloud. Since their de Broglie wavelength is negligible compared to the size of the cloud, their subsequent evolution can be analyzed by treating them as classical, pointlike particles. To this end, we employ the WKB approximation, substituting the ansatz $\chi=A(\mathbf{r},t) e^{iS(\mathbf{r},t)}$ into the Klein-Gordon equation:
	\begin{equation}
		[\nabla^\mu \nabla_{\mu}-(\mu_{\chi}^2+\lambda \phi)]\chi=0\,.
	\end{equation}
	In the leading-order eikonal limit, this procedure yields a Hamilton-Jacobi equation for the phase, which is equivalent to the on-shell condition for a classical particle. (see Appendix~\ref{app:weakfield} for details). This defines a position- and time-dependent effective mass squared for the $\chi$ particles:
	\begin{equation}\label{effmass}
		m_{\mathrm{eff}}^2=\mu_{\chi}^2+\lambda \phi(\mathbf{r},t)\,.
	\end{equation}
	The particles relevant for ring formation are those produced and trapped near the cloud's peak at $r=2r_0$ in the equatorial plane. Since this is deep within the weak-gravity region ($r\sim 2r_{0}\gg r_{g}$), we are justified in adopting the weak-field metric to describe the spacetime:
	\begin{equation}
		\dd s^2=-(1+2\Phi_N(r))\dd t^2+(1-2\Phi_N(r))\dd{\mathbf{x}}^2\,,
	\end{equation}
	where $\Phi_N(r)=-GM/r$ is the Newtonian potential. Within this framework, we can obtain the effective single-particle Lagrangian. (see Appendix~\ref{app:weakfield} for details):
	\begin{equation}
		L=-m_{\text {eff }}(1+\Phi_N) \sqrt{1-\dot{\mathbf{r}}^{2}}\,.
	\end{equation}
	The corresponding Euler-Lagrange equation of motion is
	\begin{equation}\label{EoM}
		\frac{\dd}{\dd t}\left(m_{\mathrm{eff}}(1+\Phi_N) \gamma \dot{\mathbf{r}}\right)=-\frac{1}{\gamma}\left[(1+\Phi_N) \nabla m_{\mathrm{eff}}+m_{\mathrm{eff}} \nabla \Phi_N\right]\,,
	\end{equation}
    where $\gamma=\sqrt{1-\dot{\mathbf{r}}^2}$ is the Lorentz factor.
	The resulting equation of motion is the key to understanding the postproduction fate of the $\chi$ particles. It describes the particle's trajectory under the combined influence of the weak gravitational potential $\Phi_N$ and the nongravitational force from the gradient of the $\phi$ cloud, $\sim -\nabla m_{\mathrm{eff}}$.

    \begin{figure}[h]
		\centering
            \hspace{-0.05\textwidth}
		\includegraphics[width=0.45\textwidth]{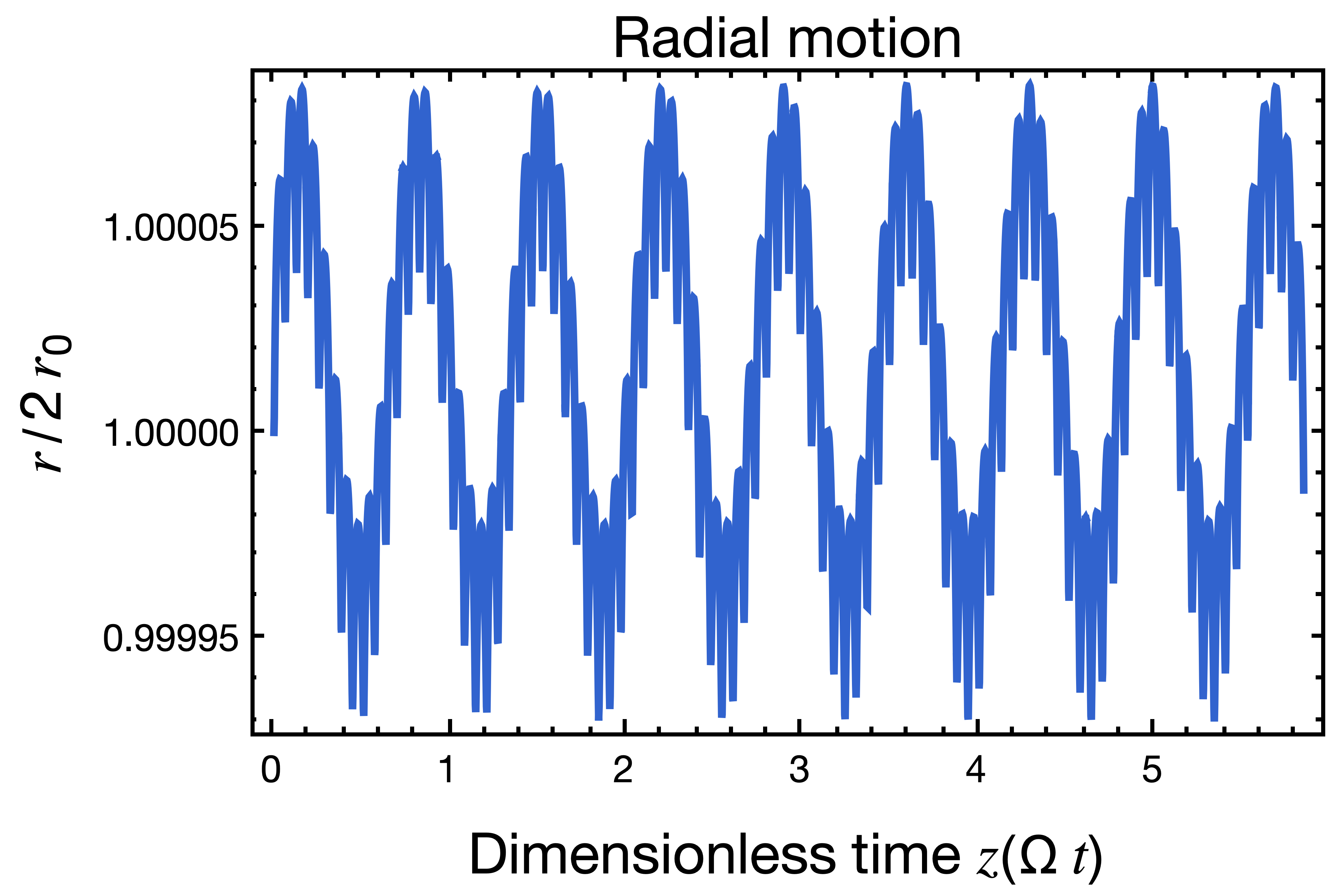}
		\caption{The radial motion of a particle in a numerically solved, stable quasicircular orbit, plotted over one orbital period with benchmark parameters ($\mu_\chi/\mu_\phi=100$, $\lambda\Phi_0/\mu_\chi^2=1$, and $\alpha=0.15$). Initial conditions are $r_i=2r_0$, $\theta_i=\pi/2$, and $\varphi_i=\pi+0.1$, modeling the particle initially produced near the resonance region with angular velocity $\dot{\varphi}_i\simeq 1.24\alpha$ and initial momenta $k\simeq 0.2\mu_\chi$. The radial coordinate $r$ oscillates with only small deviations around the initial radius $2r_0$, confirming the stability of the orbit.}
		\label{radial_motion}
    \end{figure}
	
    \begin{figure}[!h]
		\centering
		\includegraphics[width=0.5\textwidth]{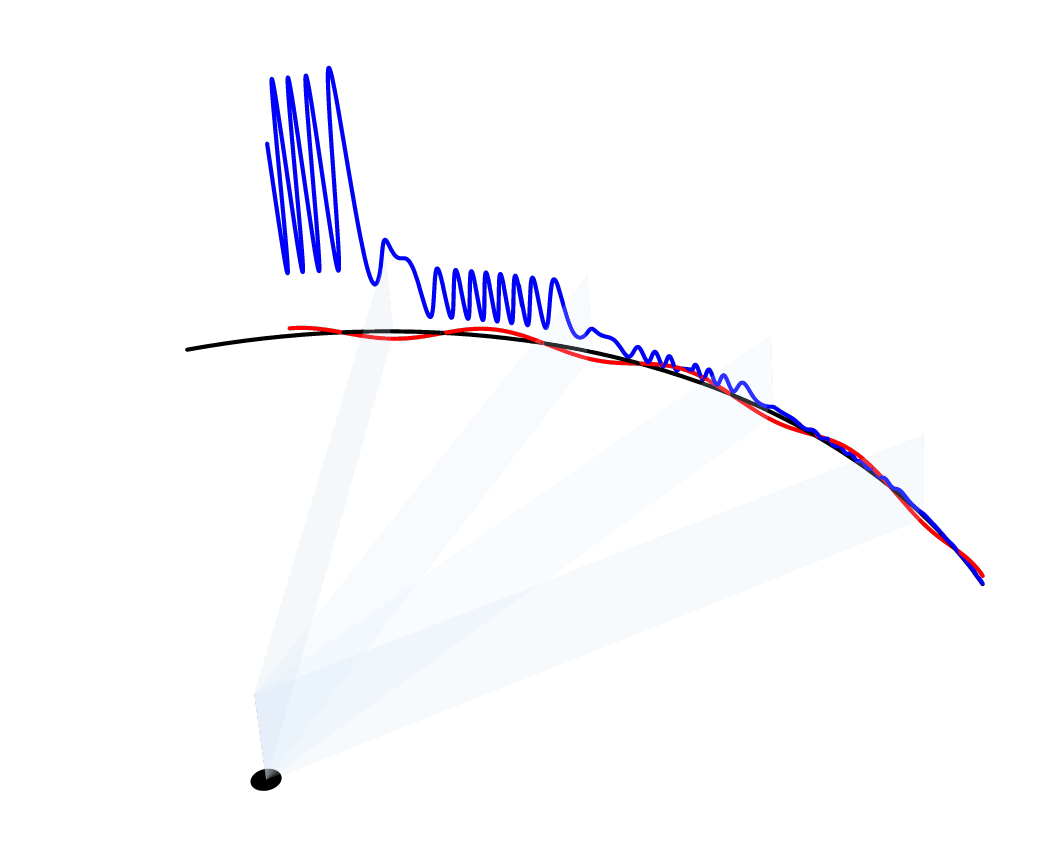}
		\caption{The illustration of the cumulative amplification mechanism for a trapped $\chi$ particle. This figure schematically shows the ``staircase" growth of a trapped mode's occupation number $\abs{n_{{k}_{\star}}}$ as it moves along its quasicircular orbit. The smooth, horizontal segment (black line) represents the particle's motion between resonant regions, where its amplitude remains constant. The sharp vertical jumps in the blue line represent brief, intense amplification of the occupation number $n_{{k}_{\star}}$ as the particle crosses resonance planes~(the gray planes), occurring each time $\phi$ oscillates to its minimum (the phase of $\phi$ is shown by the red line).}
		\label{illustration}
    \end{figure}	
    
    For most particles, the initial ``kick" from resonance places them on trajectories that are quickly destabilized by the perturbing force, causing them to either escape to infinity or be accreted by the black hole. A distinct dynamical pathway, however, emerges for only a select fraction of $k$ modes generated with initial conditions that favor a quasicircular orbit. The key lies in the spatially dependent profile of the superradiant cloud, which peaks at the toroidal region. This naturally singles out a special class of particles: those generated with low tangential momentum $k_\star\sim \alpha \mu_\chi$. This condition makes sure that the particles periodically pass through the resonance region and generate more particles on the same orbit. Over time, these selected modes become dominant, as other particles—produced initially—follow trajectories that quickly take them out of the optimal resonance region.
    
    While the nongravitational force from the cloud, $\sim -\nabla m_{\mathrm{eff}}$, acts as a force that would normally destabilize a simple Keplerian orbit, a stable quasicircular path is nevertheless possible. This is due to a crucial separation of timescales: the force from the $\phi$ field oscillates rapidly at frequency $\omega\sim \mu_\phi$, whereas the particle's orbital frequency is much lower, i.e., $\Omega\sim \alpha^2\omega\ll\omega$. Over a single orbit, the effects of this rapidly oscillating force are largely averaged out, allowing the particle to maintain its near-circular trajectory. Our numerical solutions to Eq.~\eqref{EoM} confirm that such trajectories are indeed possible, as illustrated in Fig.~\ref{radial_motion}, which shows a particle's radial coordinate remaining stable with slight oscillations. The long-term stability of these orbits is further confirmed by a formal perturbative analysis detailed in Appendix~\ref{app:stability}.
    
    This orbital configuration enables the mechanism of cumulative amplification. Particles that enter the quasicircular orbit will repeatedly traverse the resonant regions. The number of crossings per orbit is substantial, i.e., $N_{\text{crossings}} = {\mu_\phi}/{\Omega} \sim {\alpha^{-2}}$. This implies that approximately $\alpha^{-2}$ particle production events occur per orbital cycle, demonstrating that amplification efficiency in such orbits far exceeds all other orbital configurations.

    This hybrid process—classical orbital motion punctuated by bursts of amplification—is what leads to the ``staircase" growth of the particle's occupation number, as schematically depicted in Fig.~\ref{illustration}. These frequent, periodic amplifications provide a unique mechanism for particles in these select orbits to continuously extract energy from the $\phi$ cloud. In contrast, other particles on different trajectories rapidly diminish in the fraction of the total $\chi$ particles. This cumulative energy input drives the virialization of the trapped particles, causing them to settle into a quasi-steady-state configuration. The final structure, as hypothesized, is a toroidal ring located in the region of most efficient production and amplification: a circular band around $r\approx 2r_0$ in the equatorial plane. Note that our WKB-based description of a classical particle on a well-defined trajectory is valid during the orbital periods between resonance events; in this phase, particle number is conserved. The classical picture breaks down only for the brief moments when the particle traverses the resonance planes. At these instances, the process is properly described as an amplification of the $\chi$ field. Since the duration of each amplification event is negligible compared to the overall orbital period, these periodic ``kicks" do not significantly alter the long-term trajectory. Therefore, the particle's overall path remains well described by the effective classical equation of motion, Eq.~\eqref{EoM}.
    
    \begin{figure}[h]
		\centering
		\includegraphics[width=0.35\textwidth]{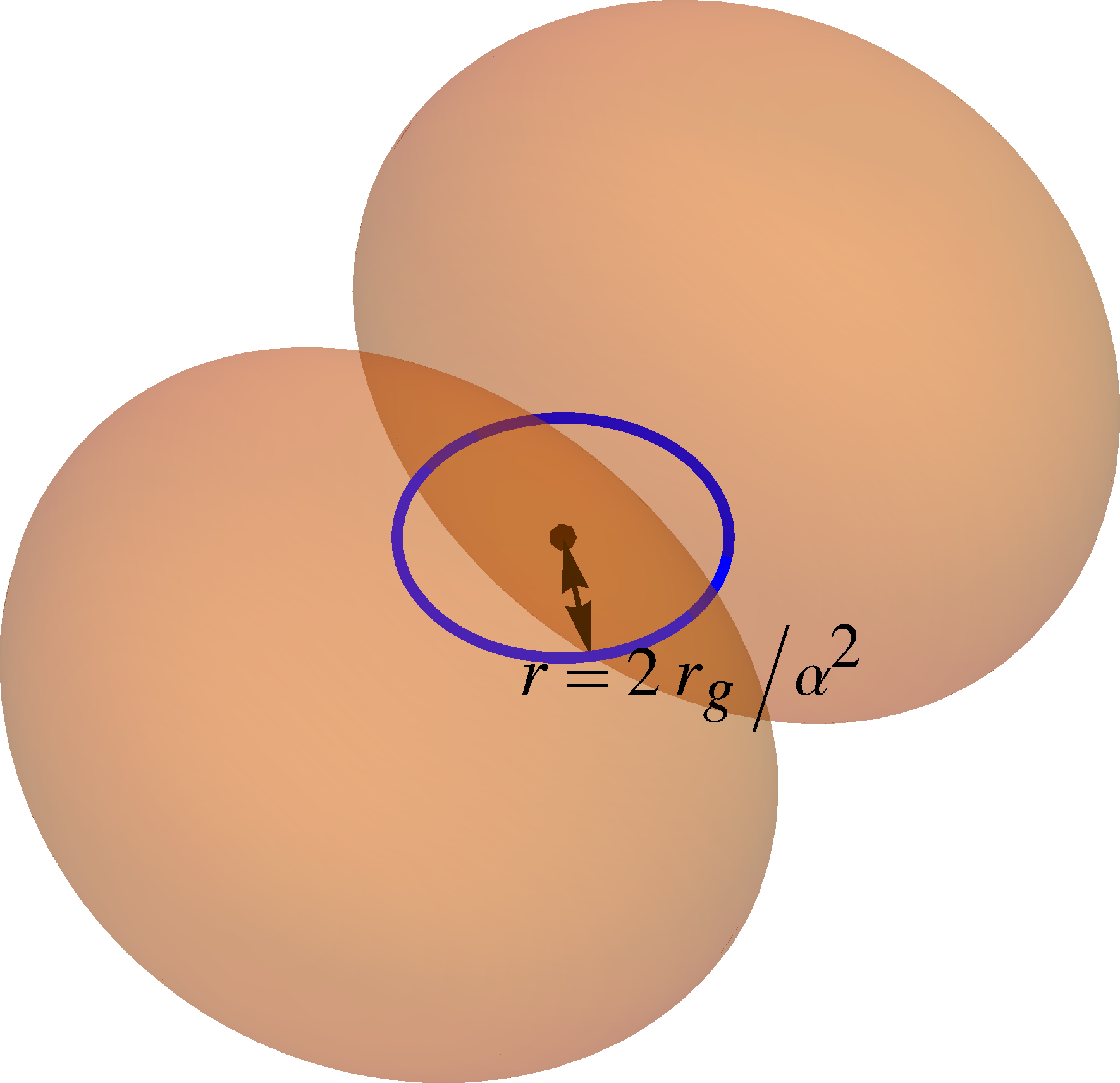}
		\caption{A schematic of the final cloud-ring configuration. The larger, transparent orange structure shows the toroidal density profile of the primary superradiant cloud of the $\phi$ axion (in the (211) state). Embedded within this cloud, near its density peak at $r=2r_0=2r_g/\alpha^2$ is the stable secondary ring composed of $\chi$ particles (blue solid ring). This illustrates the saturated final state of the system, where the primary cloud coexists with the particle ring it has created.}
		\label{cloudring}
    \end{figure}

	\subsection{Backreaction and saturation}
	
	The cumulative amplification described in the previous section provides a powerful mechanism for transferring energy from the $\phi$ cloud to the $\chi$ ring. This process, however, cannot continue indefinitely. As the $\chi$ ring grows in mass and density, its backreaction on the primary $\phi$ cloud becomes significant, eventually halting the energy transfer. In this section, we analyze this backreaction process to determine the final, saturated state of the system.

	\setlength{\tabcolsep}{5pt}
	\begin{table*}[t]
		\centering
		\caption{Benchmark parameters for the saturated cloud-ring system. We present four representative cases: two for a stellar-mass black hole ($M_{\text{BH}}=12.5 M_\odot$) and two for a supermassive black hole ($M_{\text{BH}}=2.5 \times 10^9 M_\odot$). For each, we show results for an early saturation (smaller $\Phi_{0c}$) versus a late saturation (larger $\Phi_{0c}$) scenario. The specific parameters used for these scenarios are as follows. For the stellar-mass case, we use $\alpha=0.1$, $\mu_\phi=10^{-12}\,\mathrm{eV}$, $\mu_\chi=10^{-10}\,\mathrm{eV}$ with $\lambda=10^{-40}\,\mathrm{eV}$ (for the early saturation case) and $\lambda=10^{-44}\,\mathrm{eV}$ (for the late saturation case). For the SMBH case, we use $\alpha=0.2$, $\mu_\phi=10^{-20}\,\mathrm{eV}$, $\mu_\chi=10^{-17}\,\mathrm{eV}$ with $\lambda=10^{-54}\,\mathrm{eV}$ (for the early saturation case) and $\lambda=10^{-59}\,\mathrm{eV}$ (for the late saturation case).}
		\label{tab:results}
		\begin{tabular}{|c|c|c|c|c|c|c|}
			\hline
			\text{Scenario} & $M_{\text{BH}}$ & $\Phi_{0c}$ & $M_{\text{cloud}}$ & $M_{\chi}$ & $N_\chi$ & $M_{\chi}/M_{\text{cloud}}$ \\ \hline
			Stellar BH (Early Sat.) & $12.5 M_\odot$ & $10^{11} \text{GeV}$ & $1.67 \times 10^{-9} M_\odot$ & $10^{-13} M_\odot$ & $10^{63}$ & $6 \times 10^{-5}$ \\ \hline
			Stellar BH (Late Sat.) & $12.5 M_\odot$ & $10^{15} \text{GeV}$ & $0.167 M_\odot$ & $10^{-5} M_\odot$ & $10^{71}$ & $6 \times 10^{-5}$ \\ \hline
			SMBH (Early Sat.) & $2.5 \times 10^{9} M_\odot$ & $10^{11} \text{GeV}$ & $0.021 M_\odot$ & $1.25 \times 10^{-8} M_\odot$ & $10^{75}$ & $6 \times 10^{-7}$ \\ \hline
			SMBH (Late Sat.) & $2.5 \times 10^{9} M_\odot$ & $10^{16} \text{GeV}$ & $2.08 \times 10^{8} M_\odot$ & $125 M_\odot$ & $10^{85}$ & $6 \times 10^{-7}$ \\ \hline
		\end{tabular}
	\end{table*}
	
	The energy transfer from the $\phi$ cloud to the $\chi$ ring is governed by
	\begin{align}
		\dot{M}_{\mathrm{cloud}}^{(\phi)} &= 2 \Gamma_{211} M_{\mathrm{cloud}}^{(\phi)} - \dot{M}_{\mathrm{prod}}\,, \\
		\dot{M}_{\chi} &= \dot{M}_{\mathrm{prod}}\,,
	\end{align}
	where the cloud growth rate can be obtained directly from Eq.~\eqref{eq:211} and Eq.~\eqref{cloudmass}, i.e., $2\Gamma_{211}M_{\text {cloud }}^{(\phi)}\simeq 7.75\, \tilde{a}\alpha^{5}\Phi_{0}^{2}$. $\dot{M}_{\mathrm{prod}}$ is the production rate of $\chi$ particles, which can be expressed as
	\begin{equation}\label{productionrate}
		\dot{M}_{\mathrm{prod}}\simeq \frac{\mu_{\phi}}{\sqrt{ 2 }\pi^3\mathcal{N}}\sqrt{ \lambda \Phi_{0} }\exp(4\pi\mathcal{N} \tau_{k})\,,
	\end{equation}
	where we have averaged over $\mathcal{N}$ oscillation period of $\phi$ field. $\tau_{k}\sim\mathcal{O}(0.1)$ is a numeric resonance rate. Thus, it is expected that the growth of $\phi$ cloud stops when $\dot{M}_{\rm prod}$ becomes comparable to the superradiant growth rate $2 \Gamma_{211} M_{\mathrm{cloud}}^{(\phi)}$. This indicates that the energy extraction rate from the black hole via superradiance balances the energy transfer rate to the $\chi$-field sector, establishing a steady-state regime where the cloud mass is roughly unchanged.
	
	To proceed with the subsequent backreaction estimate, we therefore make a physically motivated, simplifying assumption. We postulate that the system saturates shortly after the parametric resonance becomes efficient. 
	According to Eq.~\eqref{eq:condition}, we can approximate the final amplitude of the cloud as the value set by the resonance threshold itself:
	\begin{equation}\label{saturation}
		\Phi_{0c} \simeq \frac{\mu_\chi^2}{\lambda}\,.
	\end{equation}
	
	We can numerically validate this simplifying assumption with a self-consistency check. Once the cloud's amplitude reaches the resonance threshold, $\Phi_{0}(t) \simeq {\mu_\chi^2}/{\lambda}$, the particle production rate begins to grow exponentially, as shown in Eq.~\eqref{productionrate}. For this production rate to become large enough to balance the cloud growth rate, it typically requires a duration corresponding to $\mathcal{N}\simeq \mathcal{O}(100)$ oscillation periods. During this interval, $\Delta t=2\pi\mathcal{N}/\mu_\phi$, the amplitude $\Phi_0(t)$ grows by a factor of $\exp(2\Gamma_{211}\Delta t)$. Given that the superradiant rate is typically very small compared to the axion frequency $\mu_\phi$, this growth factor is negligible (i.e., very close to 1). For instance, for a benchmark case with $\alpha=0.2$, $\mu_\chi=10^{-10}\,\mathrm{eV}$, $\mu_\phi=10^{-12}\,\mathrm{eV}$, and $\lambda=10^{-40}\,\mathrm{eV}$, we find that saturation is reached after only $\mathcal{N}\approx 57$ oscillations, during which time the cloud's amplitude increases by a mere $10^{-4}$. Therefore, the cloud's amplitude is effectively ``frozen" at the resonance threshold value. 
	This confirms that our approximation Eq.~\eqref{saturation} is robust and provides a reasonable estimate for the final saturation amplitude.
	
	As $\chi$ production grows, the number density of the particle ring, denoted by $n_\chi$, increases, and its backreaction on the source $\phi$ cloud becomes increasingly significant, eventually saturating the growth of the ring. This dynamic can be understood by examining the equation of motion for $\phi$, which includes a source term from the produced $\chi$ particles:
	\begin{equation}\label{eq:source}
		(\partial^2-\mu_{\phi}^2)\phi=\frac{1}{2}\lambda\chi^2\,.
	\end{equation}
	Since the source term $\left\langle\chi^{2}\right\rangle$ contains rapidly oscillating terms and their frequencies are much higher than $\mu_\phi$, we can safely apply the oscillation averaged expectation value $\frac{1}{2}\lambda\ev{\chi^2}$ to estimate the source term in Eq.~\eqref{eq:source}. After this averaging, the term can be well approximated by the local number density (see Appendix~\ref{app:backreaction} for a detailed derivation):
	\begin{equation}
		\ev{\chi^2}\approx \frac{n_{\chi}}{\sqrt{ \mu_{\chi}^2+\lambda \phi(t) }}\,.
	\end{equation}
	
	The backreaction becomes significant when this new source term becomes comparable to the original mass term in the equation, i.e., $\frac{1}{2}\lambda \ev{\chi^2}\sim \mu_{\phi}^2\phi$, as treated in Ref.~\cite{Kofman:1997yn}. From this condition, we can estimate the critical number density, $n_{\chi,c}$, at which the growth of the ring saturates,
	\begin{equation}
		n_{\chi,c}\sim \frac{2\mu_{\phi}^2}{\lambda}\Phi_{0c}\sqrt{ \mu_{\chi}^2+\lambda\Phi_{0c} }\,.
	\end{equation}
	where $\Phi_{0c}$ is the saturated amplitude of the $\phi$ field in Eq.~\eqref{saturation}. Once the number density of the $\chi$ ring reaches this critical value, the energy transfer process is saturated, and the total mass of the ring stabilizes.
	
	Building on this, we can now estimate the total mass of the saturated $\chi$ ring, $M_\chi$. The total mass of the ring is found by integrating the critical number density over the volume ($M_\chi = \mu_\chi n_{\chi, c} \mathcal{V}$). We approximate the ring as a torus, estimating its volume as the product of its major circumference $2\pi r_0$ and its cross-sectional area taken as $\pi k_\star^{-1}$ with $k_\star\sim\alpha\mu_\chi$ being the characteristic momentum of the trapped particles. A detailed calculation yields
	\begin{equation}
		M_{\chi} \approx \frac{8\sqrt{2}\pi^2}{\alpha^3}\frac{\mu_{\phi}}{\mu_{\chi}^2}\Phi_{0c}^2\,.
	\end{equation}
	This shows that the final mass of the ring scales quadratically with the saturation amplitude. For comparison, the total mass of the primary $\phi$ cloud exhibits the same dependence as shown in Eq.~\eqref{cloudmass}. By taking the ratio of these two masses, the dependence on the cloud's amplitude $\Phi_{0c}$ cancels out, leading to a remarkably simple and predictive result:
	\begin{equation}
		\frac{M_{\chi}}{M_{\mathrm{cloud}}} \approx \frac{8\sqrt{2}\pi^2}{186} \left( \frac{\mu_{\phi}}{\mu_{\chi}} \right)^2 \approx 0.6 \left( \frac{\mu_{\phi}}{\mu_{\chi}} \right)^2\,.
		\label{eq:mass_ratio}
	\end{equation}
	This result implies that the final mass ratio between the secondary ring and the primary cloud is a constant, determined only by the fundamental mass hierarchy of the two axion fields. It is independent of the specific properties of the host black hole, such as its mass and spin. This provides a clear target for observational consequences of this ring-formation scenario. Ultimately, the system settles into a stable equilibrium where the primary $\phi$ cloud and the secondary $\chi$ ring coexist, as schematically depicted in Fig.~\ref{cloudring}.
	
	To illustrate the physical scales of the final cloud-ring system, we present several benchmark scenarios in Table~\ref{tab:results}. We consider two main situations: one corresponding to a stellar-mass black hole and the other to a supermassive black hole (SMBH). For each type of black hole, we explore two distinct saturation scenarios determined by the interaction strength $\lambda$. A larger value of $\lambda$ triggers resonance early in the superradiant growth, leading to a smaller saturation amplitude $\Phi_{0c}$. Conversely, a smaller $\lambda$ allows the $\phi$ cloud to grow to a much larger, more massive state before the ring production becomes efficient and halts the growth. As shown in the table, a later onset of resonance (a larger $\Phi_{0c}$) results in a significantly more massive final ring, as its mass scales with the total mass of the primary cloud.
	
	\section{Conclusion and Discussion}~\label{sec:Discussion}
 
	In this work, we have explored a novel phenomenology that emerges from a coupled axionlike particles system undergoing superradiance around a Kerr black hole. We demonstrated that the coupling term, $\frac{1}{2}\lambda\phi\chi^2$, can trigger the repeated resonant production of the heavier particles, $\chi$, from the primary superradiant cloud of a lighter scalar, $\phi$. 
	By analyzing the full dynamical sequence---from the initial growth of the $\phi$ cloud to the subsequent particle production---we identified a novel mechanism for the formation of a stable structure of $\chi$. We showed that particles produced with specific low, tangential momenta can become gravitationally trapped. Using a WKB analysis, we found that these particles follow quasistable circular orbits, allowing them to be repeatedly amplified by the oscillating $\phi$ cloud. This cumulative energy injection drives the system towards a saturated equilibrium, resulting in the formation of a stable ring of $\chi$ particles embedded within the primary cloud.
	Distinct from other studies of resonant particle production, this work takes into consideration the spatiotemporal onset of the resonance. We have shown that particle production does not occur uniformly throughout the cloud. Instead, it commences at the location where $\phi$ field's amplitude first reaches the resonant threshold---the toroidal peak of the wave function at $r = 2r_0$. It is this localized ignition of particle production that naturally singles out the circular orbit at this specific radius, which allows the $\chi$ particles to cross the resonant region repetitively, providing a robust physical reason for the ring's formation site and its subsequent stability. This crucial link between the cloud's spatial profile and the resulting particle dynamics has not been emphasized in previous literature.
    
    A natural question is whether particle-particle scattering processes could disrupt this stable ring structure after its formation. We have analyzed the most relevant channels and found their effects to be negligible. The annihilation of two heavy $\chi$ particles into a much lighter $\phi$ quanta is kinematically forbidden due to the large mass hierarchy ($\mu_\chi\gg\mu_\phi$), which in turn suppresses any subsequent scattering involving $\phi$ quanta. Furthermore, while elastic scattering $\chi\chi\to\chi\chi$ is possible, its rate is extremely low. The mean free path for these collisions scales with the coupling as $L\propto\lambda^{-2}$ and is found to be orders of magnitude larger than the ring's radius itself. Consequently, scattering events are far too infrequent to collectively disrupt the ring's stability.
	
	Furthermore, we briefly address the contrasting case of the quartic interaction, $\frac{1}{2}g^2\phi^2\chi^2$. While this coupling also triggers parametric resonance, it leads to a more violent and dissipative outcome. The extreme conditions require the interaction energy term $g^2\Phi_0^2$ to far exceed the particle's bare mass squared, $\mu_\chi^2$. Consequently, the postproduction dynamics are governed by a powerful nongravitational force proportional to $g^2\nabla^2\phi^2$, which completely dominates the particle's trajectory. Instead of settling into stable orbits, the produced particles are rapidly accelerated to relativistic energies and ejected from the system. This scenario is analogous to the fermion acceleration by the scalar cloud discussed in Ref.~\cite{Chen:2023vkq}. The final state is therefore not a stable ring but a powerful, escaping flux that acts as a dissipative channel, quenching the cloud's growth. A detailed analysis of the resonance condition and particle acceleration for this case is deferred to Appendix~\ref{app:interaction}.
	
	The final, stable cloud-ring system may also offer new phenomenological possibilities. In the presence of a binary companion, for instance, it is known that resonance transition from tidal perturbations can deplete the primary $\phi$ cloud efficiently~\cite{Baumann:2018vus,Berti:2019wnn}. Our work suggests a potential consequence of such an event: the rapid depletion of the $\phi$ cloud could alter the dynamical environment for the secondary $\chi$ ring, possibly causing it to evolve into a more diffuse, extended halo over time. The presence of this new halo of heavy particles might then leave a distinct imprint on the subsequent gravitational wave signal from the binary inspiral, providing an indirect but powerful probe of this phenomenology.

	\begin{acknowledgments}
		This work is supported in part by the National Key Research and Development Program of China Grants No. 2020YFC2201501 and No. 2021YFC2203002, in part by the National Natural Science Foundation of China Grants No. 12475067, No. 12347103, No. 12235019, No. 12075297, No. 12147103 and No.12588101, in part by the Science Research Grants from the China Manned Space Project with NO. CMS-CSST-2021-B01, in part by the Fundamental Research Funds for the Central Universities.

	\end{acknowledgments}

\appendix

\section{Analysis of Additional Interaction Terms}
\label{app:interaction}

In this Appendix, we analyze the effects of various other possible interaction terms on the dynamics of the $\chi$ field.

\textit{Linear $\chi$ interactions $\phi\chi$, $\phi^2\chi$, and $\phi^3\chi$}—These couplings induce forced, rather than parametric, resonance. For instance, For example, with a background oscillation of the $\phi$ field—$\phi(t) = \Phi_0\cos(\mu_\phi t)$—the ${\phi}^2\chi$ interaction produces an inhomogeneous equation of motion
\begin{equation}
  \ddot{\chi} + (k^2 + \mu_{\chi}^2)\chi = \sigma \Phi_{0}^2 \cos^2(\mu_{\phi} t)\,.
\end{equation}
Resonance would require the driving frequency from the $\phi$ field to match the energy of the produced $\chi$ particle, for example, $\sqrt{k^2 + \mu_{\chi}^2} \approx 2\mu_{\phi}$. Given the large mass hierarchy in our setup ($\mu_{\phi} \ll \mu_{\chi}$), this condition is kinematically forbidden, as the low-energy quanta of the $\phi$ field cannot efficiently produce a much heavier $\chi$ particle.

\textit{Cubic $\chi$ interaction term $\phi\chi^3$}—For the nonlinear interaction $V_{\mathrm{int}} \propto \kappa \phi\chi^3$, the equation of motion for $\chi$ under the oscillating background of the $\phi$ cloud becomes
\begin{equation}
  \ddot{\chi} + (k^2 + \mu_{\chi}^2)\chi + 3 \kappa \Phi_{0} \cos(\mu_{\phi} t)\,\chi^2 = 0\,.
\end{equation}
Compared to the standard linear Mathieu equation, the oscillatory term here is quadratic. Although narrow resonance at $k^2 + \mu_\chi^2 \simeq \mu_\phi^2$ is kinematically forbidden since $\mu_\chi \gg \mu_\phi$, broad resonance may still occur. However, this instability is suppressed by the nonlinearity of the $\chi^2$ term and is less efficient than our primary focus.

\textit{Quartic interaction $\phi^2\chi^2$}—The quartic coupling $V_{\rm int}=\frac{1}{2}g^2\phi^2\chi^2$ leads to parametric resonance. The equation of motion for a $\chi_k$ mode is a Mathieu-type equation:
\begin{equation}
	\ddot{\chi}_{k}+\left(\mu_{\chi}^{2}+k^{2}+g^{2} \Phi_{0}^{2} \cos^{2} (\mu_{\phi} t-\varphi)\right) \chi_{k}=0\,.
\end{equation}
The resonance band for this equation occurs when the condition $k_\star^2+\mu_\chi^2 \approx g\Phi_{0}\mu_\phi$ is met. For this resonance to be broad and efficient enough to compete with superradiance, a more stringent condition on the dimensionless coupling parameter is required, namely $g^2\Phi_0^2/\mu_\chi^2 \sim (\mu_\chi/\mu_\phi)^2$.

\begin{figure}[!h]
	\centering
	\includegraphics[width=0.35\textwidth]{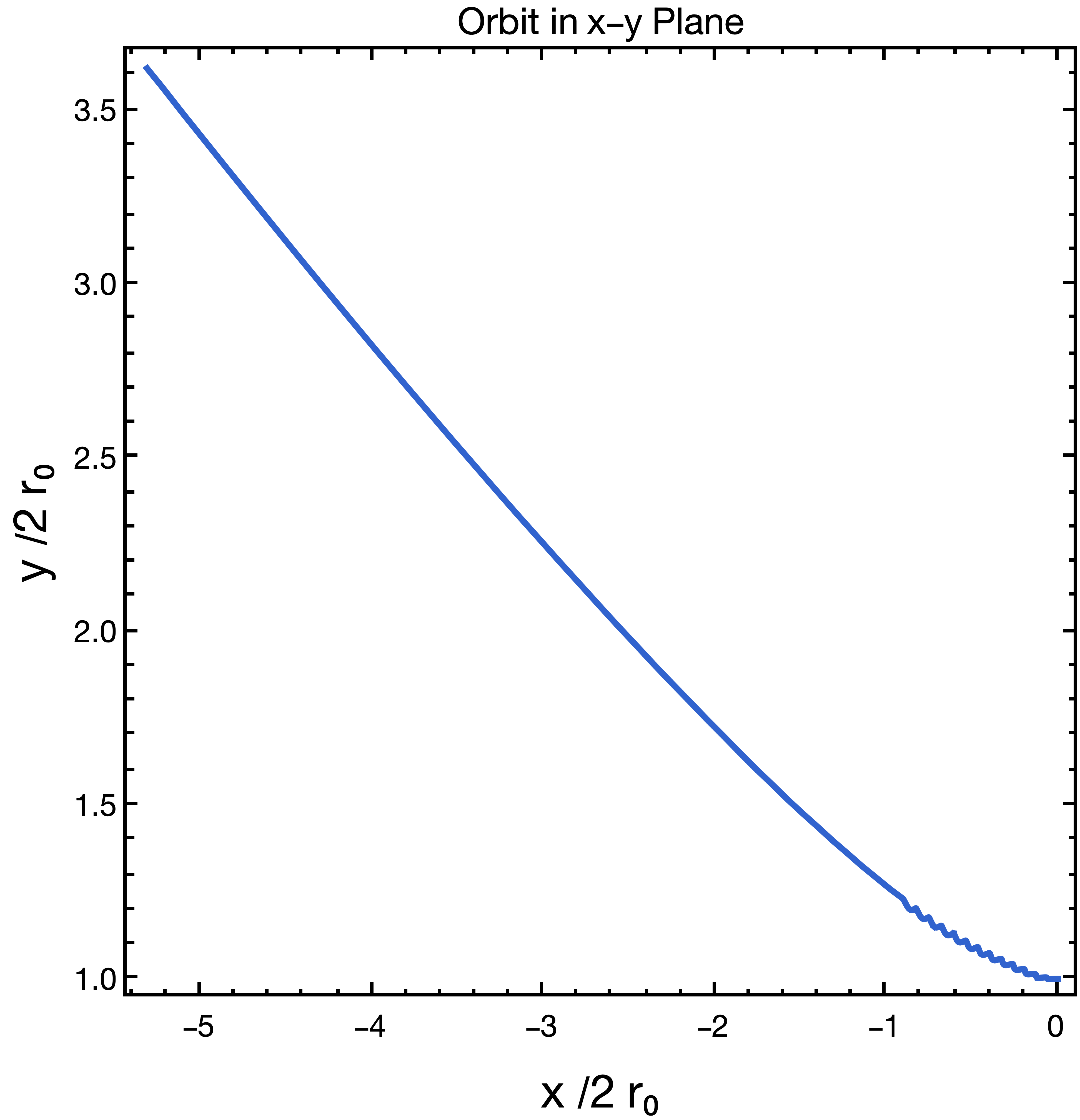}
	\caption{Particle trajectory in the equatorial plane for the quartic interaction scenario. The parameters used for this simulation are $\mu_\chi/\mu_\phi=100$, $g^2\Phi_0^2/\mu_\chi^2=10^4$, and $\alpha=0.1$.}
	\label{fourpointtrajectory}
\end{figure}
\begin{figure}[!h]
	\centering
	\includegraphics[width=0.4\textwidth]{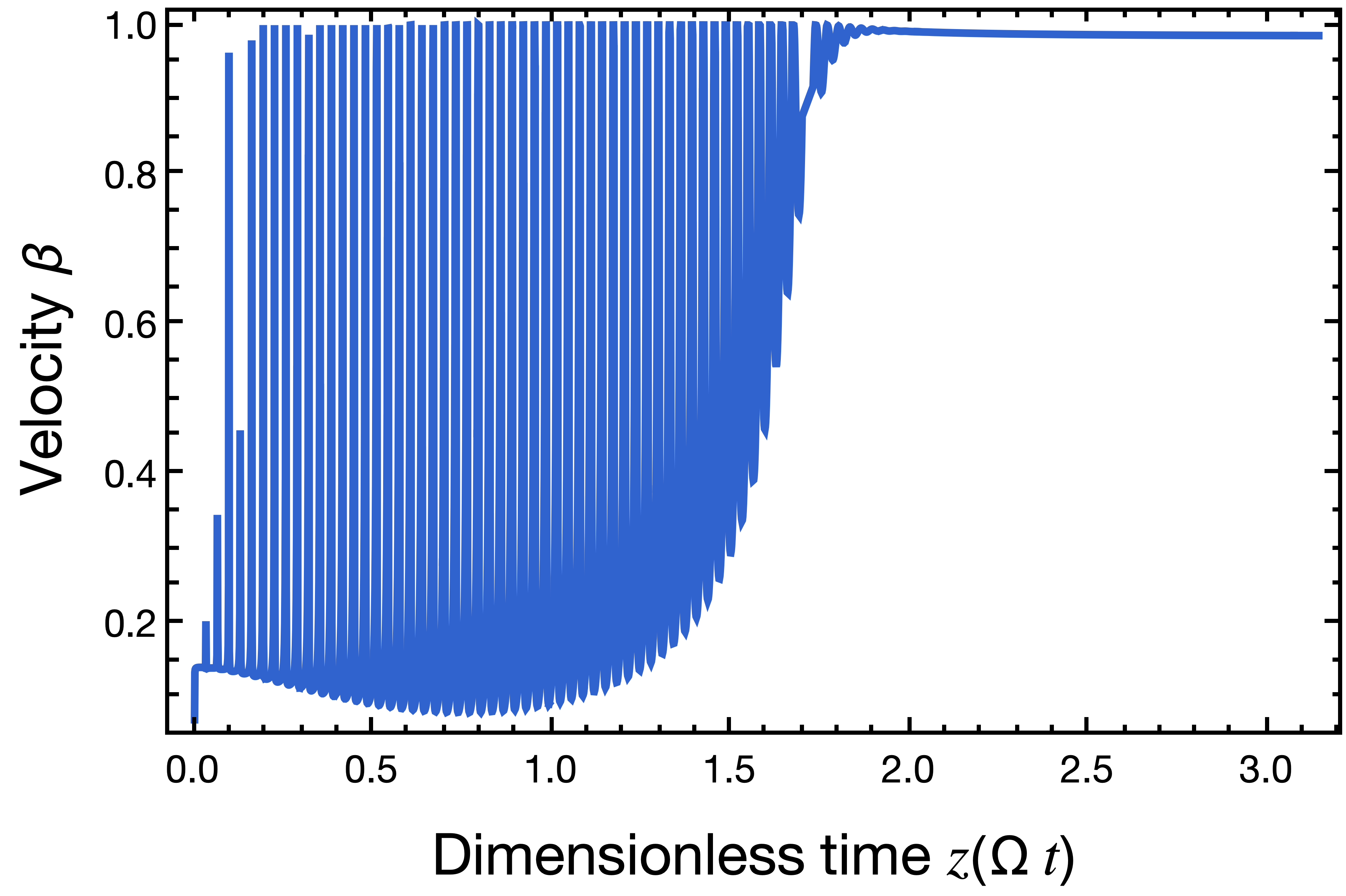}
	\caption{Evolution of particle velocity over time for the quartic interaction scenario. The dimensionless time is defined by $z=\Omega t$ (see Appendix~\ref{app:stability}). The parameters are the same as above.}
	\label{fourpointLorentz}
\end{figure}

This implies that the interaction energy term must be much larger than the particle's bare mass squared, $g^2\Phi_0^2 \gg \mu_\chi^2$. In this regime, the dynamics is dominated by the nongravitational force proportional to $g^2\nabla(\phi^2)$. Our numerical simulations and stability analysis (see Appendix~\ref{app:stability}) of the particle trajectories both confirm that this dominant force renders the orbits unstable and rapidly accelerates the produced $\chi$ particles to relativistic energies. This process is illustrated in a benchmark simulation in Fig.~\ref{fourpointtrajectory} and Fig.~\ref{fourpointLorentz}. The particle is quickly ejected on an escaping trajectory, and its Lorentz factor, after an initial period of rapid oscillation, grows to a large, constant value, confirming the particle becomes ultra-relativistic. This result is consistent with the findings of Ref.~\cite{Chen:2023vkq}, which demonstrated through a similar analysis that a strong gradient force from a scalar cloud can accelerate fermions to energies on the order of $g\Phi_0$. The final state is therefore an escaping flux rather than a stable, bound ring.

\section{WKB and Weak-Field Approximation}
\label{app:weakfield}

Here, we provide a detailed derivation of the equation of motion for the $\chi$ particle in the presence of the external $\phi$ field and gravity. We first adopt the WKB approximation, which is valid in the short-wavelength limit $k \gg \mu_{\phi}\alpha$. We thus write the $\chi$ field as a wave with a slowly varying amplitude $A(\mathbf{r}, t)$ and a rapidly varying phase $S(\mathbf{r}, t)$:
\begin{equation}
\chi_{k}(\mathbf{r}, t)=A(\mathbf{r}, t) e^{i S(\mathbf{r}, t)}\,.\label{WKB}
\end{equation}

Substituting this ansatz into the Klein-Gordon equation for $\chi$ in curved spacetime, and keeping only the leading-order terms in the derivative expansion, yields the Hamilton-Jacobi equation:
\begin{equation}
g^{\mu \nu}\partial_{\mu}S\partial_{\nu}S=-m_{\mathrm{eff}}^2\,.
\end{equation}

This equation can be interpreted geometrically by defining the particle's four-momentum as the gradient of the phase, $k_{\mu}=\partial_{\mu}S$. With this definition, the Hamilton-Jacobi equation is equivalent to the on-shell condition for a relativistic particle:
\begin{equation}
g^{\mu \nu}k_{\mu}k_{\nu}=-m_{\mathrm{eff}}^2\,,
\end{equation}
where the particle propagates as if it has a position- and time-dependent effective mass squared, given by Eq.~\eqref{effmass}.

To find the equation of motion, we differentiate both sides with respect to an affine parameter $\tau$. The left-hand side becomes:
\begin{equation}
k^{\alpha} \nabla_{\alpha} \bigl(k^{\sigma} k_{\sigma}\bigr) = 2\,k_{\sigma} \bigl(k^{\alpha} \nabla_{\alpha} k^{\sigma}\bigr) = 2\,k_{\sigma} K^{\sigma}\,,
\end{equation}
where we define $K^{\mu} \equiv k^{\alpha} \nabla_{\alpha} k^{\mu}$, which represents the deviation from standard geodesic motion (for which $K^\mu=0$). The derivative of the right-hand side is:
\begin{equation}
-2m_{\mathrm{eff}}\,k^{\alpha} \nabla_{\alpha} m_{\mathrm{eff}}\,.
\end{equation}

Equating the two expressions and dividing by two yields a condition on the geodesic deviation vector:
\begin{equation}
k_{\sigma} K^{\sigma} = -m_{\mathrm{eff}}\,k^{\alpha} \nabla_{\alpha} m_{\mathrm{eff}}\,.
\end{equation}
This shows that the projection of the deviation vector $K^\sigma$ along the four-momentum is related to the rate of change of the effective mass. Since the deviation is sourced by the gradient of the scalar field, it is natural to propose that $K^\mu$ is proportional to the gradient of the effective mass itself: $K^{\mu} = C(x) \,\nabla^{\mu} m_{\mathrm{eff}}$. Substituting this ansatz into the previous equation allows us to solve for the proportionality factor: $C(x)=-m_{\mathrm{eff}}$.

Having determined the form of the deviation vector, we arrive at the modified geodesic equation:
\begin{equation}
k^{\alpha} \nabla_{\alpha} k^{\mu} = -m_{\mathrm{eff}}\,\nabla^{\mu} m_{\mathrm{eff}}.
\end{equation}
This can be written more explicitly as
\begin{equation}
\frac{\mathrm{d}k^{\alpha}}{\mathrm{d}\tau} + \Gamma^{\alpha}_{\rho\sigma} k^{\rho} k^{\sigma} = -\tfrac12 \,\nabla^{\alpha} m_{\mathrm{eff}}^2\,.
\end{equation}

The term on the right-hand side acts as a nongravitational force, arising from the gradient of the $\phi$ field, which deflects the particle from a standard geodesic path, consistent with similar analyses for fermions in Ref.~\cite{Chen:2023vkq}.

Crucially, particles with these initial conditions remain confined to the weak-gravity region of the spacetime ($r\sim 2r_0 \gg r_g$). This allows us to adopt the weak-field (Newtonian) metric to describe their dynamics\footnote{The use of this non-rotating metric is justified because the leading-order effect of the black hole's spin---the frame-dragging term $g_{t\varphi}\sim GM\tilde{a}\sin^2\theta/r$---is suppressed relative to the Newtonian potential $\Phi_N$ at the large radius of the cloud, $r \sim r_0 = r_g/\alpha^2$. Furthermore, the superradiance process itself could efficiently extract angular momentum, reducing the black hole's spin and further diminishing this effect.}:
\begin{equation}
  \mathrm{d}s^2 = -\bigl(1 + 2\Phi_N(r)\bigr)\,\mathrm{d}t^2 + \bigl(1 - 2\Phi_N(r)\bigr)\,\mathrm{d}\mathbf{x}^2\,,
\end{equation}
where $\Phi_N(r) = -GM/r$ is the Newtonian gravitational potential.  In this limit, the d'Alembertian operator acting on $\chi$ decomposes as
\begin{equation}
  \Box \chi = I_1 + I_2 + I_3,
\end{equation}
with
\begin{equation}
	I_1 = \eta^{\mu \nu} \partial_{\mu} \partial_{\nu} \chi\,, I_2 = \delta g^{\mu \nu} \partial_{\mu} \partial_{\nu} \chi\,, I_3 = \partial_{\mu} \delta g^{\mu \nu} \partial_{\nu} \chi\,.
\end{equation}

Since temporal derivatives dominate over spatial ones, the leading-order contribution to $I_2$ is
\begin{equation}
  I_2 \approx 2\,\Phi_N(r)\,\partial_{t}^2 \chi\,,
\end{equation}
and because $\delta g^{\mu \nu}$ is time-independent, we can write
\begin{equation}
  I_3 = -2\,\nabla\bigl[\Phi_N(r)\,\nabla\chi\bigr]\,.
\end{equation}

Collecting terms, the full equation of motion for $\chi$ in the weak-field approximation becomes
\begin{equation}
	\partial_{t}^{2} \chi-\nabla^{2} \chi+\left(m_{\chi}^{2}+\lambda \phi^{2}\right) \chi=2\Phi_N \partial_{t}^2\chi-2\nabla(\Phi_N \nabla \chi)\,,\label{weakfieldEOM}
\end{equation}
where the right-hand side represents the correction from weak gravity. Substituting the WKB ansatz Eq.\eqref{WKB} into Eq.\eqref{weakfieldEOM}, we have the modified energy-momentum relation
\begin{equation}
	\omega_{k}^{2}= \mathbf{k}^{2}+\frac{m_{\mathrm{eff}}^{2}(\mathbf{r}, t)}{1-2 \Phi_N(r)}\approx \mathbf{k}^{2}+(1+2 \Phi_N(r)) m_{\mathrm{eff}}^{2}(\mathbf{r}, t)\,,
\end{equation}
where $\omega_{k}=\partial_{t}S$, $k_{i}=\partial_{i}S$. The motion of the wave packet center is determined by the group velocity, corresponding to the phase gradient $v^i=\frac{\partial\omega_{k}}{\partial k_{i}}=\frac{k^i}{\omega_{k}}$.

The Hamiltonian is
\begin{equation}
	H=[\mathbf{k}^{2}+(1+2 \Phi_N(r)) m_{\mathrm{eff}}^{2}(\mathbf{r}, t)]^{1/2}\,.
\end{equation}

The Lagrangian is
\begin{equation}
\begin{aligned}
	L&=-m_{\mathrm{eff}}(\mathbf{r}, t) \sqrt{(1+2 \Phi_N)\left(1-\dot{\mathbf{r}}^{2}\right)}\\
	&\approx -m_{\text {eff }}(1+\Phi_N) \sqrt{1-\dot{\mathbf{r}}^{2}}\,,
\end{aligned}
\end{equation}
where in the second line we have approximated $\sqrt{1+2\Phi_N}\approx 1+\Phi_N$.

\section{Orbital Stability Analysis}
\label{app:stability}
In this Appendix, we provide a stability analysis for the quasicircular particle orbits that are essential for ring formation. Our starting point is the full, nonlinear equation of motion~Eq.\eqref{EoM}, which is derived from the Lagrangian in the weak-field limit. We present here in its spatial component form:

\begin{widetext}
\begin{equation}
\begin{aligned}
&\left(1-\Phi_N\right)m_{\mathrm{eff}}\frac{d}{d t}\left(\gamma \dot{r}\right)-m_{\mathrm{eff}}\left(1-\Phi_N\right) \gamma r\left(\dot{\theta}^{2}+\sin ^{2} \theta \dot{\varphi}^{2}\right)=-m_{\mathrm{eff}}\gamma \frac{\dot{r}^2}{r} \Phi_N-\left( 1-\Phi_N \right)\gamma\dot{m}_{\mathrm{eff}}\dot{r}-\frac{1}{\gamma}\left(\frac{\partial m_{\mathrm{eff}}}{\partial r}\left(1-\Phi_N\right)+m_{\mathrm{eff}}\Phi_N r\right)\,,\\
&\left(1-\Phi_N\right)m_{\mathrm{eff}}\frac{d}{d t}\left(\gamma r^{2} \dot{\theta}\right)-m_{\mathrm{eff}}\left(1-\Phi_N\right) \gamma r^{2} \sin \theta \cos \theta \dot{\varphi}^{2}=-m_{\mathrm{eff}}\gamma r\dot{r}\dot{\theta} \Phi_N-\left( 1-\Phi_N \right)\gamma\dot{m}_{\mathrm{eff}}r^2\dot{\theta}-\frac{1}{\gamma} \frac{\partial m_{\mathrm{eff}}}{\partial \theta}\left(1-\Phi_N\right)\,,\\
&\left(1-\Phi_N\right)m_{\mathrm{eff}}\frac{d}{d t}\left(\gamma r^{2} \sin^{2} \theta \dot{\varphi}\right)=-m_{\mathrm{eff}}\gamma \sin^2\theta r\dot{r}\dot{\varphi}\Phi_N-\left( 1-\Phi_N \right)\gamma\dot{m}_{\mathrm{eff}}r^2\sin^2\theta\dot{\varphi}-\frac{1}{\gamma} \frac{\partial m_{\mathrm{eff}}}{\partial \varphi}\left(1-\Phi_N\right)\,,
\end{aligned}
\end{equation}
\end{widetext}
where $\gamma=\sqrt{1-\dot{\mathbf{r}}^2}$, $\Phi_N=-GM/r$, and $m_{\mathrm{eff}}^2=\mu_\chi^2+\lambda\Phi_{0\star}r/(2r_0)e^{1-r/2r_0}\sin\theta\cos(\varphi-\mu_\phi t)$.

To analyze the orbit stability, we linearize these equations around a circular reference path in the equatorial plane, defined by $r(t)=2r_0$, $\theta(t)=\pi/2$ and $\varphi = C t$, where $C$ is some constant angular velocity. We introduce small perturbations to this state: ${\mathbf{r}}(t)=\mathbf{r}_0(t)+\delta\mathbf{r}(t)$. After a lengthy but straightforward process of linearization and converting to dimensionless coordinates $x=r/(2r_0)$ and $z=\Omega t$, we obtain a set of coupled linear ordinary differential equations for the perturbations. Notably, the vertical motion $\delta\theta$ decouples from the equatorial plane motion ($\delta x$, $\delta\varphi$). The linearized equations are

\begin{widetext}
\begin{equation}
    \kappa \cos ((g-h)z)\left(-2\beta_0^{2}\delta\theta^{\prime\prime} + (1 - 3\beta_0^{2}h^{2})\delta\theta\right) - 2\beta_0^{2}\left(\delta\theta^{\prime\prime} + h^{2}\delta\theta\right) + \beta_0^{2}\kappa(g-h)\sin((g-h)z)\delta\theta^{\prime} = 0\,,
\end{equation}

\begin{equation}
    \begin{aligned}
        &\Bigl[ \kappa \bigl( 3\beta_0^{2} - \beta_0^{6} h^{4} + \beta_0^{4} h^{2}(h^{2}-4) + 1 \bigr) \cos((g-h)z) + 2\beta_0^{2} \bigl( (1-3\beta_0^{2}) h^{2} + 2 \bigr) \Bigr]\delta x \\
        &- \beta_0^{2} \Bigl[ \kappa(h^{2}-1)(\beta_0^{2} h^{2}-1) \sin((g-h)z) \delta \varphi - \kappa (\beta_0^{2}-1) (\beta_0^{2} h^{2}-1) (g-h) \sin((g-h)z) \delta x^{\prime} \\
        &+ 2(\kappa \cos((g-h)z) + 1) \Bigl( h\bigl(\beta_0^{2}(h^{2}+1)-2\bigr) \delta \varphi^{\prime} + (\beta_0^{2}-1)(\beta_0^{2} h^{2}-1) \delta x^{\prime \prime} \Bigr) \Bigr]=0 \\
        & \kappa (\beta_0^{2}-1) \delta \varphi(z) (\beta_0^{2} h^{2}-1) (\beta_0^{2} g h-1) \bigl( \kappa(\cos(2(g-h)z) + 3) + 4 \cos((g-h)z) \bigr) \\
        &+ 4\beta_0^{2} (\kappa \cos((g-h)z) + 1) \Bigl[ (\beta_0^{2}-1) \bigl( \kappa(g-h) \sin((g-h)z) \delta \varphi^{\prime}(z)- 2 (\kappa \cos((g-h)z) + 1) \delta \varphi^{\prime \prime} \bigr) \\
        &- 2h \bigl( \beta_0^{2}(h^{2}+1)-2 \bigr) (\kappa \cos((g-h)z) + 1) \delta x + \kappa \bigl( h(\beta_0^{2}(g h^{2}+g-2h)-2g+h) + 1 \bigr) \sin((g-h)z) \delta x \Bigr]=0\,.
    \end{aligned}
\end{equation}
\end{widetext}
The parameters are defined as follows: $g=\mu_\chi/\mu_\phi$, $\kappa=\lambda\Phi_{0\star}/\mu_\chi^2$, $\Omega=\sqrt{GM/(2r_0)^3}=\alpha/2\sqrt{2}r_0$, $\beta_0=\Omega (2r_0)=\alpha/\sqrt{2}$, $h=C/\Omega$.

To determine the stability of this system of linear equations with periodic coefficients, we employ Floquet theory. The procedure involves converting the second-order perturbation equations into a first-order matrix system of the form $\dot{\mathbf{Y}}(z) = \mathbf{A}(z) \mathbf{Y}(z)$, where $\mathbf{Y}(z)$ is the state vector of perturbations and their derivatives (e.g., $\mathbf{Y}_\theta(z) = [\delta\theta(z), \delta\dot{\theta}(z)]^T$) and $\mathbf{A}(z)$ is the periodic coefficient matrix.

For a linear system with a periodic matrix $\mathbf{A}(z)$ of period $T$, Floquet theory states that the long-term stability is determined by the monodromy matrix, $\mathbf{M}$. This constant matrix maps any initial perturbation state $\mathbf{Y}(0)$ to its state after one full period $T = 2\pi/|h - g|$:
\begin{equation}
\mathbf{Y}(T) = \mathbf{M}\mathbf{Y}(0)\,.
\end{equation}
The matrix $\mathbf{M}$ is computed by numerically integrating the system $\dot{\mathbf{Y}} = \mathbf{A}(z)\mathbf{Y}$ over one period with a set of standard basis vectors as initial conditions. The stability is determined by the eigenvalues, $\rho_i$, of $\mathbf{M}$, known as the Floquet multipliers. An orbit is stable only if all of its Floquet multipliers have a magnitude less than or equal to one, $|\rho_i| \le 1$.

We compute the monodromy matrices for the vertical and planar motions, $\mathbf{M}_\theta$ and $\mathbf{M}_{x\varphi}$, numerically. Our analysis confirms that there are significant regions of parameter space where all Floquet multipliers satisfy the stability condition. This analytical result depicts a particle maintaining a stable, localized trajectory in the equatorial plane, and is consistent with by direct numerical integration of the full nonlinear equations of motion presented in the main text (see Fig.~\ref{radial_motion}). 

Conversely, we have applied the same Floquet analysis to the particle trajectories produced by the quartic interaction. In that case, we find that the resulting orbits are generally unstable, with Floquet multipliers satisfying $|\rho_i| > 1$. This confirms that particles are not trapped and will escape to infinity, a conclusion that is also consistent with our direct numerical solutions showing particle acceleration and ejection (see Fig.~\ref{fourpointtrajectory} and Fig.~\ref{fourpointLorentz}).

\section{Derivation of the Averaged Backreaction Term}
\label{app:backreaction}

In this Appendix, we detail the derivation of the time-averaged approximation for the backreaction term $\langle\chi^2\rangle$ used in the main text.

The analysis begins with the formal expression for the expectation value of the squared field operator. In the presence of particle production, this can be expressed in terms of the Bogoliubov coefficients, $\alpha_k$ and $\beta_k$, for each momentum mode $k$. The expectation value contains a contribution from the produced particles (with occupation number $n_k = |\beta_k|^2$) and an oscillatory term~\cite{Kofman:1997yn}:
\begin{equation}
	\langle\chi^2\rangle = \int\frac{\dd[3]k}{(2\pi)^3}\frac{1}{\omega_k(t)} \left( n_k + \operatorname{Re}\left[\alpha_{k} \beta_{k}^{*} e^{-2 i \int_{0}^{t} \omega_{k}(t') dt'}\right] \right)\,,
\end{equation}
where we have omitted the vacuum fluctuation terms, as they can be renormalized into the mass of $\phi$.

The crucial part of the expression is the oscillatory term. Its time evolution is governed by the phase, which depends on the time-dependent frequency $\omega_k(t)$. In the regime of efficient production, where $\lambda\Phi_0 \gtrsim \mu_\chi^2$, the frequency is approximately $\omega_k \sim \sqrt{\mu_\chi^2+\lambda\phi}\gg \mu_\phi$, the phase in this  expression is equal to $\cos\left( \frac{4\sqrt{ \lambda \Phi }}{\mu_{\phi}}\cos(\mu_{\phi}t) \right)$ plus some small correction. The expression is thus simplified to
\begin{equation}
\left\langle\chi^{2}\right\rangle \approx \left[{1+C \cos \left( \frac{4 \sqrt{ \lambda \Phi_{0} }}{\mu_{\phi}}\cos \mu_{\phi} t \right)}\right] \int_{0}^{\infty} \frac{\dd[3] k}{(2\pi)^3\omega_{k}} n_{k}\,,	
\end{equation}
where $C$ is some constant. Consequently, the phase term $e^{-2i \int \omega_k dt'}$ oscillates at a very high frequency compared to the evolution of the background field $\phi(t)$ itself, which occurs at the low frequency $\mu_\phi$. When $\langle\chi^2\rangle$ acts as a source in the equation of motion for $\phi$, the contribution from this rapidly oscillating term is negligible~\cite{Kofman:1997yn}. Therefore, for modeling the backreaction on the evolution of the $\phi$ cloud, the expectation value is dominated by the particle number term: $\overline{\langle\chi^2\rangle} \approx \int \frac{d^3k}{(2\pi)^3} \frac{n_k}{\omega_k(t)}$.

By defining the total number density as $n_\chi = \int \frac{d^3k}{(2\pi)^3} n_k$, we arrive at the simplified expression used in the main text:
\begin{equation}
	\langle\chi^2\rangle \approx \frac{n_\chi}{{\omega}_k(t)} \approx \frac{n_\chi}{\sqrt{\mu_\chi^2+\lambda\phi(t)}}\,.
\end{equation}
It should be noted that while the integral for $n_\chi$ is formally over all modes, the long-term, stable particle number density is dominated by the trapped particles forming the ring. Thus, $n_\chi$ in this context effectively represents the number density of the saturated ring.

	%
	\bibliographystyle{apsrev4-1}
	\bibliography{ring}

\end{document}